\documentclass{article}
\usepackage{spconf,amsmath,graphicx}
\usepackage{amssymb}
\usepackage{color}
\usepackage{caption}
\usepackage{subcaption}

\def\0{{\mathbf 0}}
\def\1{{\mathbf 1}}

\def\x{{\mathbf x}}

\def\A{{\mathbf A}}

\def\I{{\mathbf I}}

\def\R{{\mathbf R}}
\def\S{{\mathbf S}}

\def\cF{{\mathcal F}}
\def\cG{{\mathcal G}}

\def\bPhi{{\boldsymbol \Phi}}
\def\bPsi{{\boldsymbol \Psi}}

\title{Volumetric Attribute Compression for 3D Point Clouds \\ using Feedforward Network with Geometric Attention}
\name{
    Tam Thuc Do$^{\star,\dag}$\thanks{This work was supported in part by a Google student research project.},
    Philip A. Chou$^\star$, 
    Gene Cheung$^\dag$\thanks{Gene Cheung acknowledges the additional support of the NSERC grants RGPIN-2019-06271,  RGPAS-2019-00110.}
}
\address{
  $^\star$Google Research, Seattle, Washington, USA\\ 
  $^\dag$York University, Toronto, Canada
}
\begin{document}
\ninept
\maketitle
\begin{abstract}
We study 3D point cloud attribute compression using a volumetric approach: given a target volumetric attribute function $f : \mathbb{R}^3 \rightarrow \mathbb{R}$, we quantize and encode parameter vector $\theta$ that characterizes $f$ at the encoder, for reconstruction $f_{\hat{\theta}}(\x)$ at known 3D points $\x$'s at the decoder. 
Extending a previous work Region Adaptive Hierarchical Transform (RAHT) that employs piecewise constant functions to span a nested sequence of function spaces, we propose a feedforward linear network that implements higher-order B-spline bases spanning function spaces without eigen-decomposition. 
Feedforward network architecture means that the system is amenable to end-to-end neural learning. 
The key to our network is space-varying  convolution, similar to a graph operator, whose weights are computed from the known 3D geometry for normalization.
We show that the number of layers in the normalization at the encoder is equivalent to the number of terms in a matrix inverse Taylor series.
Experimental results on real-world 3D point clouds show up to 2-3 dB gain over RAHT in energy compaction and 20-30\% bitrate reduction. 
\end{abstract}
\vspace*{-0.5ex}
\section{Introduction}
\label{sec:intro}
We perform attribute compression for 3D point clouds.  A 3D point cloud is a set of points $\{(\mathbf{x}_i,\mathbf{y}_i)\}$ with positions $\mathbf{x}_i\in\mathbb{R}^3$ and corresponding attributes $\mathbf{y}_i\in\mathbb{R}^r$.  Point cloud {\em geometry} and {\em attribute} compression respectively concern compression of the point positions and attributes --- with or without loss.  Although the geometry and attribute compression may be done jointly (e.g., \cite{AlexiouTE:20}), the dominant approach in both recent research \nocite{ZhangFL:14,ThanouCF:16,QueirozC:16,QueirozC:17b,CohenTV:16,ChouKK:20,KrivokucaCK:20}\cite{ZhangFL:14}--\cite{KrivokucaCK:20}  as well as in the MPEG geometry-based point cloud compression (G-PCC) standard \nocite{SchwarzEtAl:18,GraziosiEtAl:20,JangEtAl:19,GPCC:20}\cite{SchwarzEtAl:18}--\cite{GPCC:20} is to compress the geometry first and then to compress the attributes conditioned on the geometry.\footnote{If the geometry compression is lossy, then the original attributes must first be {\em transferred} to the reconstructed geometry before being compressed.}  Thus, even if the geometry is lossy, the problem of attribute compression reduces to the case where the decoded geometry is known at both the attribute encoder and at the attribute decoder.  We make this assumption here.

There are many ways to perform attribute compression.  
We take a \textit{volumetric approach}.  
We say a real vector-valued function $f:\mathbb{R}^d\rightarrow\mathbb{R}^r$ is {\em volumetric} if $d=3$ (or {\em hyper-volumetric} if $d>3$).  
Examples of volumetric functions are absorption densities, signed distances to nearest surfaces, radiance fields, and light fields.\footnote{Radiance or light fields, which depend not only on a position $\mathbf{x}\in\mathbb{R}^3$ but also on a direction $(\theta,\phi)$ on the unit sphere may be represented either by a hyper-volumetric function of $(x,y,z,\theta,\phi)$ that yields a vector of attributes $\mathbf{a}$ or a volumetric function of $(x,y,z)$ that yields a vector of parameters (e.g., spherical harmonics) of an attribute-valued function on the sphere $\mathbf{a}(\theta,\phi)$.}
In the volumetric approach to point cloud attribute compression, a function $f_\theta$ is fitted to the attributes $\mathbf{y}_i$ at the positions $\mathbf{x}_i$ (assumed known) of the point cloud.  
Then $f_\theta$ is compressed and transmitted by compressing and transmitting quantized $\hat{\theta}$.  
Finally, the volumetric function $f_{\hat\theta}$ is reproduced at the decoder, and queried at the points $\mathbf{x}_i$ to obtain reconstructed attributes $\hat{\mathbf{y}}_i=f_{\hat\theta}(\mathbf{x}_i)$.  It may also be queried any point $\mathbf{x}\in\mathbb{R}^3$, to achieve infinite spatial zoom, for example.

The volumetric approach to point cloud attribute compression is uncommon in the literature.  
However, \cite{ChouKK:20} showed that volumetric attribute compression is at the core of MPEG G-PCC, which uses the Region Adaptive Hierarchical Transform (RAHT) \cite{QueirozC:16,SandriCKQ:19}.  
RAHT implicitly projects any volumetric function $f$ that interpolates attributes $\mathbf{y}_i$ at positions $\mathbf{x}_i$ onto a nested sequence of function spaces $\mathcal{F}_0^{(1)}\subseteq\cdots\subseteq\mathcal{F}_L^{(1)}$ at different levels of detail $\ell=0,\ldots,L$. 
In particular, $\mathcal{F}_\ell^{(1)}$ is the space of functions that are piecewise constant over cubes of width $2^{L-\ell}$ voxels, where $2^L$ is the size in voxels of a cube containing the point cloud.  That is, $\mathcal{F}_\ell^{(1)}$ is the space of (3D) B-splines of order 1 at scale $2^{L-\ell}$.  Further, \cite{ChouKK:20} showed how to improve RAHT by instead using the spaces $\mathcal{F}_\ell^{(p)}$ of B-splines of order $p>1$, which we call $p$-th order RAHT, or RAHT($p$), in this paper.  However, projecting $f$ onto $\mathcal{F}_\ell^{(p)}$ for $p>1$ required computing large eigen-decompositions in \cite{ChouKK:20}.

\begin{figure}
    \centering
    \includegraphics[width=0.48\linewidth]{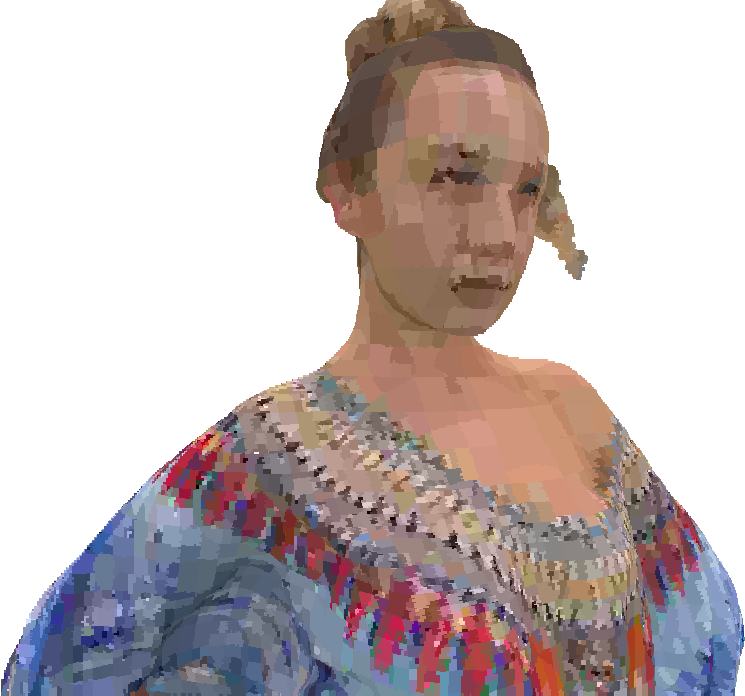}
    \includegraphics[width=0.48\linewidth]{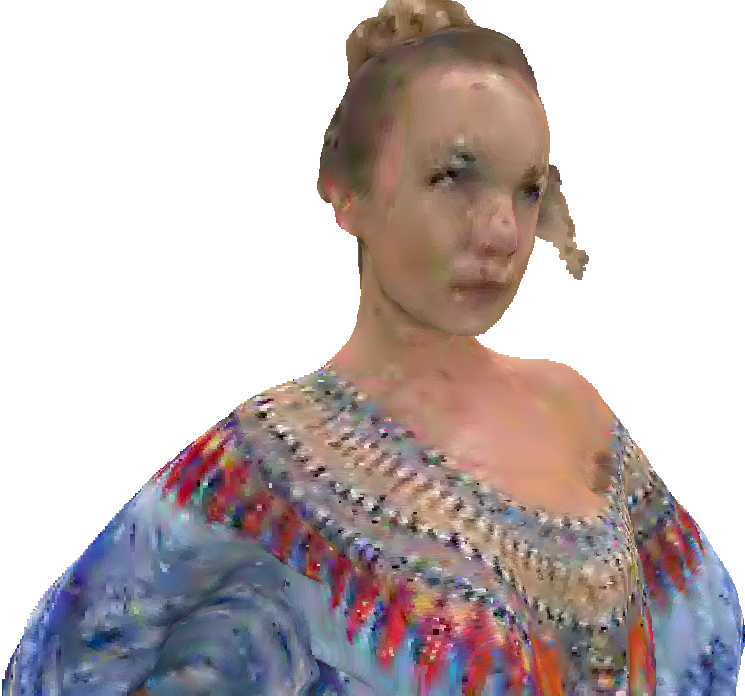}
    \caption{Reconstructed PC for RAHT($p=2$) at 0.22 bpp (right) and RAHT($p=1$) and 0.20 bpp  (left)}
    \label{fig:recon_PC_samebpp}
    \vspace*{-3ex}
\end{figure}

Meanwhile, following the success of learned neural networks for image compression \nocite{ToOMHwViMi16,BaLaSi16a,ToViJoHwMi17,BaLaSi17,Ba18,BaMiSiHwJo18,MiBaTo18,BalleEtAl:20,mentzer2020high,hu2021learning}\cite{ToOMHwViMi16}--\cite{hu2021learning}, a number of works have successfully improved on the point cloud geometry compression in MPEG G-PCC using learned neural networks \nocite{YanSLLLL:19,quach2019learning,GuardaRP:19a,GuardaRP:19b,guarda2020deep,tang2020deep,Quach2020ImprovedDP}\cite{YanSLLLL:19}--\cite{Quach2020ImprovedDP}.  
A few works have tried using learned neural networks for point cloud attribute compression as well \cite{AlexiouTE:20}\nocite{ShengLLXLW:21,IsikCHJT:21,WangM:22}\cite{ShengLLXLW:21}--\cite{WangM:22}.  Other works have used neural networks for volumetric attribute compression more generally \nocite{BirdBSC:21,isik2021neural,TakikawaETMMJF22}\cite{BirdBSC:21}--\cite{TakikawaETMMJF22}.  However, on the task of point cloud attribute compression, the learned neural methods have yet to outperform MPEG G-PCC.  The reason may be traced to the fundamental difficulty with point clouds that they are sparse, irregular, and changing.  The usual dense space-invariant convolutions used by neural networks for image compression as well as point cloud geometry compression become sparse convolutions when the attributes are known at only specific positions.  Sparse convolutions (convolutions that treat missing points as having attributes equal to zero) should be normalized, but until now it has not been clear how.

In this paper, we take a step towards learning networks for point cloud attribute compression by showing how to implement RAHT($p$) as a feedforward linear network, \textit{without eigen-decompositions}.  
The backbone convolutions of our feedforward network (analogous to the downsampling convolutions and upsampling transposed convolutions in a UNet \cite{RonnebergerFB:15}) are sparse with space-invariant kernels.  The ``ribs'' of our feedforward network (analogous to the skip connections in a UNet) are normalized with space-varying convolutions.  The space-varying convolutions can be considered a graph operator with a generalized graph Laplacian \cite{cheung18}.  
We show that the weights of this operator can themselves be computed from the geometry using sparse convolutions of weights on the points with space-invariant kernels.  This is the way that the attribute transform is conditioned on the geometry, and that the geometry is used to normalize the results of the sparse convolutions.  The normalization may be interpreted as an attention mechanism, corresponding to the ``Region Adaptivity'' in RAHT.  We show that the number of layers in the normalizations at the encoder is equivalent to the number of terms in a Taylor series of a matrix inverse, or to the number of steps in the unrolling of a gradient descent optimization to compute the coefficients needed for decoding.  In our experimental results section, we study both the energy compaction and bit rate of our feedforward network, and compare to RAHT, showing up to 2--3 dB gain in energy compaction and 20--30\% reduction in bitrate.  
Although there are no nonlinearities and no learning currently in our feedforward network, we believe that this is an important first step to understanding and training a feedforward neural network for point cloud attribute compression.


\vspace*{-0.5ex}
\section{Coding Framework}
\label{sec:framework}

Assume scalar attributes. The extension to vector attributes is straightforward. Denote by $f:\mathbb{R}^3\rightarrow\mathbb{R}$ a volumetric function defining an attribute field over $\mathbb{R}^3$. 
Denote by $\{(\mathbf{x}_i,y_i)\}$ a point cloud with attributes $y_i=f(\mathbf{x}_i)$, where $\x_i \in \mathbb{R}^3$ is point $i$'s 3D coordinate. 
To code the attributes given the positions, RAHT($p$) projects $f$ onto a nested sequence of function spaces $\mathcal{F}_0^{(p)}\subseteq\cdots\subseteq\mathcal{F}_L^{(p)}$. 
Each function space $\mathcal{F}_\ell^{(p)}$ is a parametric family of functions spanned by a set of basis functions $\{\phi_{\ell,\mathbf{n}}^{(p)}:\mathbf{n}\in\mathcal{N}_\ell\}$, namely the volumetric B-spline basis functions $\phi_{\ell,\mathbf{n}}^{(p)}(\mathbf{x})=\phi_{0,\mathbf{0}}^{(p)}(2^\ell\mathbf{x}-\mathbf{n})$ of order $p$ and scale $\ell$ for offsets $\mathbf{n}\in\mathcal{N}_\ell\subset\mathbb{Z}^3$.  
Specifically, for $p=1$, $\phi_{0,\mathbf{0}}^{(p)}(\mathbf{x})$ is the product of ``box'' functions $1_{[0,1]}(x)$ over components $x$ of $\mathbf{x}$, while for $p=2$, it is the product of ``hat'' functions $1_{[-1,1]}(x)(1-|x|)$ over components $x$ of $\mathbf{x}$.  
Thus, RAHT(1) projects $f$ onto the space $\mathcal{F}_\ell^{(1)}$ of functions that are \textit{piecewise constant} over blocks of width $2^{-\ell}$, while RAHT(2) projects $f$ onto the space $\mathcal{F}_\ell^{(2)}$ of continuous functions that are \textit{piecewise trilinear} over blocks of width $2^{-\ell}$. 
In any case, it is clear that $f_\ell\in\cF_{\ell}$ implies $f_\ell\in\cF_{\ell+1}$.  
Hence, the spaces are nested.  
In the sequel, we will sometimes omit the order $(p)$ for ease of notation.

Any $f_\ell \in \mathcal{F}_\ell$  can be expressed as a linear combination of basis functions $\phi_{\ell,\mathbf{n}}$:
\vspace{-2ex}
\begin{equation} 
f_\ell(\mathbf{x}) = \sum_{\mathbf{n}\in\mathcal{N}_\ell} F_{\ell, \mathbf{n}} \phi_{\ell, \mathbf{n}}(\mathbf{x}) ,
\vspace{-0.75ex}
\end{equation}
which may also be written as $f_\ell=\bPhi_\ell F_\ell$, where $\bPhi_\ell=[\phi_{\ell,\mathbf{n}}]$ is the row-vector of basis functions $\phi_{\ell,\mathbf{n}}$, and $F_\ell=[F_{\ell,\mathbf{n}}]$ is the column-vector of coefficients $F_{\ell,\mathbf{n}}$, each of length $N_\ell=|\mathcal{N}_\ell|$.

RAHT projects $f$ on $\mathcal{F}_\ell$ in the Hilbert space of (equivalence classes of) functions with inner product $\langle f,g\rangle=\sum_if(\mathbf{x}_i)g(\mathbf{x}_i)$. 
Thus, the projection $f_\ell^*$ of $f$ on $\mathcal{F}_\ell$, denoted $f_\ell^*=f\circ\mathcal{F}_\ell$, minimizes the squared error $||f-f_\ell||^2=\langle f-f_\ell,f-f_\ell\rangle=\sum_i(y_i-f_\ell(\mathbf{x}_i))^2$, which is the squared error of interest for attribute coding.
Further, the coefficients $F_{\ell}^*$ that represent the function $f_\ell^* = f\circ\mathcal{F}_\ell$ in the basis $\bPhi_\ell$ --- called the \textit{low-pass} coefficients --- can be calculated via the \textit{normal} equation:
\vspace{-1ex}
\begin{equation} \label{eq:normal_equation_F}
F_{\ell}^* = (\bPhi_{\ell}^{\top} \bPhi_{\ell})^{-1} \bPhi_{\ell}^{\top} f .
\vspace{-1ex}
\end{equation}
Here, $\bPhi^\top_{\ell} f$ denotes the length-$N_\ell$ vector of inner products $[\langle\phi_{\ell,\mathbf{n}},f\rangle]$, and $\bPhi_{\ell}^{\top} \bPhi_{\ell}$ denotes the $N_\ell\times N_\ell$ \textit{Gram matrix} of inner products $[\langle\phi_{\ell,\mathbf{n}},\phi_{\ell,\mathbf{m}}\rangle]$.
Morover, as all the projections $f_\ell^*$ satisfy the Pythagorean Theorem \cite[Sec.~3.3]{Luenberger:69}, it can be shown \cite{ChouKK:20} that $f_\ell^*=f_{\ell'}^*\circ\mathcal{F}_\ell$ for all $\ell'\geq\ell$, and that all the {\em residual} functions
\vspace{-1ex}
\begin{equation}
    g_\ell^* = f_{\ell+1}^* - f_\ell^*
\vspace{-1ex}
\end{equation}
lie in the subspace $\mathcal{G}_\ell$ that is the orthogonal complement to $\mathcal{F}_\ell$ in $\mathcal{F}_{\ell+1}$. 
Thus,
\vspace{-1ex}
\begin{eqnarray}
\cF_L & = &\cF_0 \oplus \cG_0 \oplus \cdots \oplus \cG_\ell \oplus \cdots \cG_{L-1}   \;\;\; \mbox{and}
\label{eq:sum_of_subspaces} \\
f_L^* & = & f_0^* + g_0^* + \cdots + g_\ell^* + \cdots + g_{L-1}^* .
\vspace{-2ex}
\label{eqn:fg0_gL}
\end{eqnarray}
In our coding framework, we assume that the level of detail $L$ is high enough that $f\in\mathcal{F}_L$.  
Thus, we take $f_L^*=f$, and code $f$ by coding $f_0^*$, $g_0^*,\ldots,g_{L-1}^*$.  
To code these functions, we represent them as coefficients in a basis, which are quantized and entropy coded.  Next we show how to compute these coefficients efficiently.

\vspace*{-0.5ex}
\subsection{Low-pass coefficients}

Assume that there is a sufficiently fine resolution $L$ such that the $i$th basis function $\phi_{L,\mathbf{n}_i}$ is 1 on $\mathbf{x}_i$ and 0 on $\mathbf{x}_j$ for $j\neq i$, namely $\phi_{L,\mathbf{n}_i}(\mathbf{x}_j)=\delta_{ij}$.  
Then, we have $\bPhi_L^\top \bPhi_L = \I$.
Next, denote by $\A_{\ell}=[a_{ij}^l]$ a $N_{\ell} \times N_{\ell + 1}$ matrix whose $i$th row expresses the $i$th basis function $\phi_{\ell,\mathbf{n}_i}$ of $\mathcal{F}_\ell$ in terms of the basis functions $\phi_{\ell+1,\mathbf{m}_j}$ of $\mathcal{F}_{\ell+1}$, namely
\vspace{-1ex}
\begin{equation}
\phi_{\ell,\mathbf{n}_i} = \sum_j a_{ij}^\ell \phi_{\ell+1,\mathbf{m}_j}.
\vspace{-1ex}
\end{equation}
Note that for $\mathbf{d}_{ij}=\mathbf{m}_j-2\mathbf{n}_i$,
\vspace{-1ex}
\begin{equation}
a_{ij}^\ell = \left\{\begin{array}{ll}
1 & \mbox{if}~~ \mathbf{d}_{ij} \in \{0, 1\}^3 \\ 
0 & \mbox{otherwise}
\end{array}\right.
\label{eq:RAHT1}
\vspace*{-2ex}
\end{equation}
for $p=1$ and
\vspace*{-2ex}
\begin{equation}
a_{ij}^\ell = \left\{\begin{array}{ll}
2^{- \| \mathbf{d}_{ij} \|_1} & \mbox{if} ~~ \mathbf{d}_{ij} \in \{-1, 0, 1\}^3 \\ 
0 & \mbox{otherwise}
\end{array}\right.
\label{eq:RAHT2}
\vspace*{-1ex}
\end{equation}
for $p=2$.
In vector form, this can be written
\vspace*{-1ex}
\begin{equation}
\bPhi_\ell = \bPhi_{\ell+1} \A_\ell^\top ,
\label{eqn:Phi_ell_in_terms_of_Phi_ell1}
\vspace*{-1ex}
\end{equation}
where $\mathbf{A}_\ell$ is an ordinary (i.e., space-invariant) but {\em sparse} convolutional matrix with $2\times$ spatial downsampling, and $\mathbf{A}_\ell^\top$ is the corresponding transpose convolutional matrix with $2\times$ upsampling.  Using these, we can compute matrices of inner products of basis functions from level $\ell+1$ to level $\ell$:
\begin{eqnarray}
\bPhi_\ell^\top \bPhi_\ell & = & \A_\ell\bPhi_{\ell+1}^\top \bPhi_{\ell+1} \A_\ell^\top 
\label{eqn:PhiTPhi} \\
\bPhi_{\ell+1}^\top \bPhi_\ell & = & \bPhi_{\ell+1}^\top \bPhi_{\ell+1} \A_\ell^\top 
\label{eqn:Phi1TPhi} \\
\bPhi_\ell^\top \bPhi_{\ell+1} & = & \A_\ell \bPhi_{\ell+1}^\top \bPhi_{\ell+1} . 
\label{eqn:PhiTPhi1}
\end{eqnarray}
From these expressions, it is possible to see that the matrix $\bPhi_\ell^\top \bPhi_\ell $ is also a sparse (tri-diagonal) convolutional matrix, though space-varying, corresponding to the edge weights of a graph with 27-neighbor connectivity (including self-loops).

Thus, beginning at the highest level of detail $L$, where $\phi_{L,\mathbf{n}_i}(\mathbf{x}_j)=\delta_{ij}$ and $\bPhi_L^\top\bPhi_L=\I$, we have $\bPhi_L^\top f=[y_i]$ and hence (from \eqref{eq:normal_equation_F}) $F_L^*=(\bPhi_L^\top\bPhi_L)^{-1}\bPhi_L^\top f=[y_i]$.  
Moving to $\ell<L$, we have
\begin{align}
F_\ell^* & = (\bPhi_\ell^\top \bPhi_\ell)^{-1} \bPhi_\ell^\top f_{\ell+1}^* \\
& = (\bPhi_\ell^\top \bPhi_\ell)^{-1} \bPhi_\ell^\top \bPhi_{l+1} F_{\ell+1}^* 
\label{eqn:Fl} \\
& = (\bPhi_\ell^\top \bPhi_\ell)^{-1} \A_\ell(\bPhi_{\ell+1}^\top \bPhi_{\ell+1}) F_{\ell+1}^*.
\label{eqn:F_ell_star_from_F_ell1_star}
\end{align}
Letting $\tilde{F}_\ell^* = (\bPhi_{\ell}^\top \bPhi_{\ell}) F_{\ell}^*$, which we call un-normalized low-pass coefficients,
we can easily calculate $\tilde{F}_\ell^*$ for all levels $\ell$ as
\vspace{-1ex}
\begin{equation}
\tilde{F}_\ell^* = \A_\ell \tilde{F}_{\ell+1}^* .
\label{eqn:analysis_lowpass}
\vspace{-1ex}
\end{equation}
This is an ordinary sparse convolution.
Finally, $F_\ell^*$ can be computed:
\vspace{-1ex}
\begin{equation}
    F_\ell^* = (\bPhi_{\ell}^\top \bPhi_{\ell})^{-1} \tilde{F}_\ell^*.
    \label{eqn:normalization}
\vspace{-1ex}
\end{equation}
We will see later how to compute this using a feedforward network.




\vspace*{-0.5ex}
\subsection{High-pass coefficients}

To code the residual functions $g_0^*,\ldots,g_{L-1}^*$, they must be represented in a basis.  Since $g_\ell^*\in\mathcal{G}_\ell\subset\mathcal{F}_{\ell+1}$, one possible basis in which to represent $g_\ell^*$ is the basis $\bPhi_{\ell+1}$ for $\mathcal{F}_{\ell+1}$.  
In this basis, the coefficients for $g_\ell^*=f_{\ell+1}^*-f_\ell^*=\bPhi_{\ell+1}F_{\ell+1}^*-\bPhi_\ell F_\ell^*$ are given by the normal equation (\ref{eq:normal_equation_F}) and expression (\ref{eqn:Phi1TPhi}):
\vspace{-0.5ex}
\begin{equation}
    (\bPhi_{\ell+1}^\top\bPhi_{\ell+1})^{-1}\bPhi_{\ell+1}^\top g_\ell^* = F_{\ell+1}^*-\A_\ell^\top F_\ell^* \stackrel{\Delta}{=} \delta F_{\ell+1}^* .
    \label{eqn:delta_F}
\vspace{-0.5ex}
\end{equation}
However, the number of coefficients in this representation is $N_{\ell+1}$, whereas the dimension of $\mathcal{G}_\ell$ is only $N_{\ell+1}-N_\ell$.  
Coding $g_\ell^*$ using this representation is thus called \textbf{overcomplete (or non-critical) residual coding}.

Alternatively, $g_\ell^*$ can be represented in a basis $\bPsi_\ell$ for $\mathcal{G}_\ell$.  
Denote by $G_\ell^*$ a column-vector of coefficients corresponding to the row-vector of basis functions $\bPsi_\ell$, both of length $N_{\ell+1}-N_\ell$, such that $g_\ell^*=\bPsi_\ell G_\ell^*$.  The coefficients $G_\ell^*$ --- here called {\em high-pass} coefficients --- can be calculated via the normal equation
\vspace{-0.5ex}
\begin{equation} 
G_{\ell}^* = (\bPsi_{\ell}^{\top} \bPsi_{\ell})^{-1} \bPsi_{\ell}^{\top} g_\ell^* .
\label{eq:normal_equation_G}
\vspace{-0.5ex}
\end{equation}
Though there are many choices of basis $\bPsi_\ell$ for $\mathcal{G}_\ell$, we know that $\mathcal{G}_\ell$ is orthogonal to $\mathcal{F}_\ell$ and hence $\bPhi_{\ell}^\top \bPsi_{\ell} = \mathbf{0}$. 
Further, since $\mathcal{G}_\ell\subset\mathcal{F}_{\ell+1}$, $\bPsi_\ell$ can be written as a linear combination of basis functions $\bPhi_{\ell+1}$:
\vspace{-0.5ex}
\begin{equation}
\bPsi_{\ell} = \bPhi_{\ell+1} \mathbf{Z}_\ell^\top .
\label{eq:contruct_psi}
\vspace{-0.5ex}
\end{equation}
Hence, $\mathbf{0} = \bPhi_{\ell}^\top \bPsi_{\ell} =  \A_\ell \bPhi_{\ell+1}^\top \bPhi_{\ell+1} \mathbf{Z}_\ell^\top = \A_\ell \tilde{\mathbf{Z}}_\ell^\top$, where $\mathbf{Z}_\ell^\top = (\bPhi_{\ell+1}^\top \bPhi_{\ell+1})^{-1} \Tilde{\mathbf{Z}}_\ell^\top$.
Thus, the columns of $\tilde{\mathbf{Z}}_\ell^\top$ span the null space of $\A_\ell$.  
One possible choice is
\vspace{-1ex}
\begin{equation}
\tilde{\mathbf{Z}}_\ell^\top = \left[\begin{array}{c}
    -(\A_\ell^a)^{-1}\A_\ell^{b} \\ 
    \I^b
\end{array}\right] \label{eq:define_ZT}
\vspace{-1ex}
\end{equation}
where we split $\A_\ell$ into two sub-matrices of size $N_\ell \times N_\ell$ and $N_\ell \times (N_{\ell + 1} - N_\ell)$:
\vspace{-1ex}
\begin{equation}
\A_\ell = \left[\begin{array}{cc}
    \A_\ell^a &\A_\ell^b 
\end{array}\right] .
\end{equation}
Coding $g_\ell^*$ using the representation $G_\ell^*$ in (\ref{eq:normal_equation_G}) is called \textbf{orthonormal (or critical) residual coding}. Sec.~\ref{sec:results} compares compression performance of orthonormal and overcomplete residual coding.



\vspace*{-0.5ex}
\subsection{Orthonormalization}

Direct scalar quantization of the coefficients $F_\ell^*$ and $G_\ell^*$ do not yield good rate-distortion performance, because their basis functions $\bPhi_\ell$ and $\bPsi_\ell$, while orthogonal to each other (i.e., $\bPhi_\ell^\top \bPsi_\ell=\mathbf{0}$), are not orthogonal to themselves (i.e., $\bPhi_\ell^\top \bPhi_\ell \neq \I$, and $\bPsi_\ell^\top \bPsi_\ell \neq \I$). 
Hence, we \textit{orthonormalize} them as follows:
\vspace{-1ex}
\begin{align}
\bar{\bPhi}_\ell &= \bPhi_\ell \R_\ell = \bPhi_\ell (\bPhi_\ell^\top \bPhi_\ell)^{-\frac{1}{2}} \\
\bar{\bPsi}_\ell &= \bPsi_\ell \S_\ell = \bPsi_\ell  (\bPsi_\ell^\top \bPsi_\ell)^{-\frac{1}{2}} .
\end{align}
In this way, $\bar{\bPhi}_\ell^\top \bar{\bPhi}_\ell = \I$ and $\bar{\bPsi}_\ell^\top \bar{\bPsi}_\ell = \I$. 
The coefficients corresponding to the orthonormal basis functions $\bar{\bPhi}_\ell$ and $\bar{\bPsi}_\ell$ can now be calculated as:
\vspace{-1ex}
\begin{align}
    \bar{F}_{\ell}^* &= \bar{\bPhi}_{\ell}^{\top} f =(\bPhi_{\ell}^{\top} \bPhi_{\ell})^{-\frac{1}{2}} \bPhi_{\ell}^{\top} f = (\bPhi_{\ell}^{\top} \bPhi_{\ell})^{\frac{1}{2}} F_\ell^* \label{eqn:orthor_normal_equation_F} \\
    \bar{G}_{\ell}^* &= \bar{\bPsi}_{\ell}^{\top} g =(\bPsi_{\ell}^{\top} \bPsi_{\ell})^{-\frac{1}{2}} \bPsi_{\ell}^{\top} g = (\bPsi_{\ell}^{\top} \bPsi_{\ell})^{\frac{1}{2}} G_\ell^*.
\label{eq:orthor_normal_equation_G}
\end{align}

\vspace*{-1.5ex}
\section{Feedforward Network Implementation}
\label{sec:implement}

\begin{figure}
\centering
    \includegraphics[
    height=0.3\textheight, trim=1.0in 0.125in 5.125in 0.25in,clip]{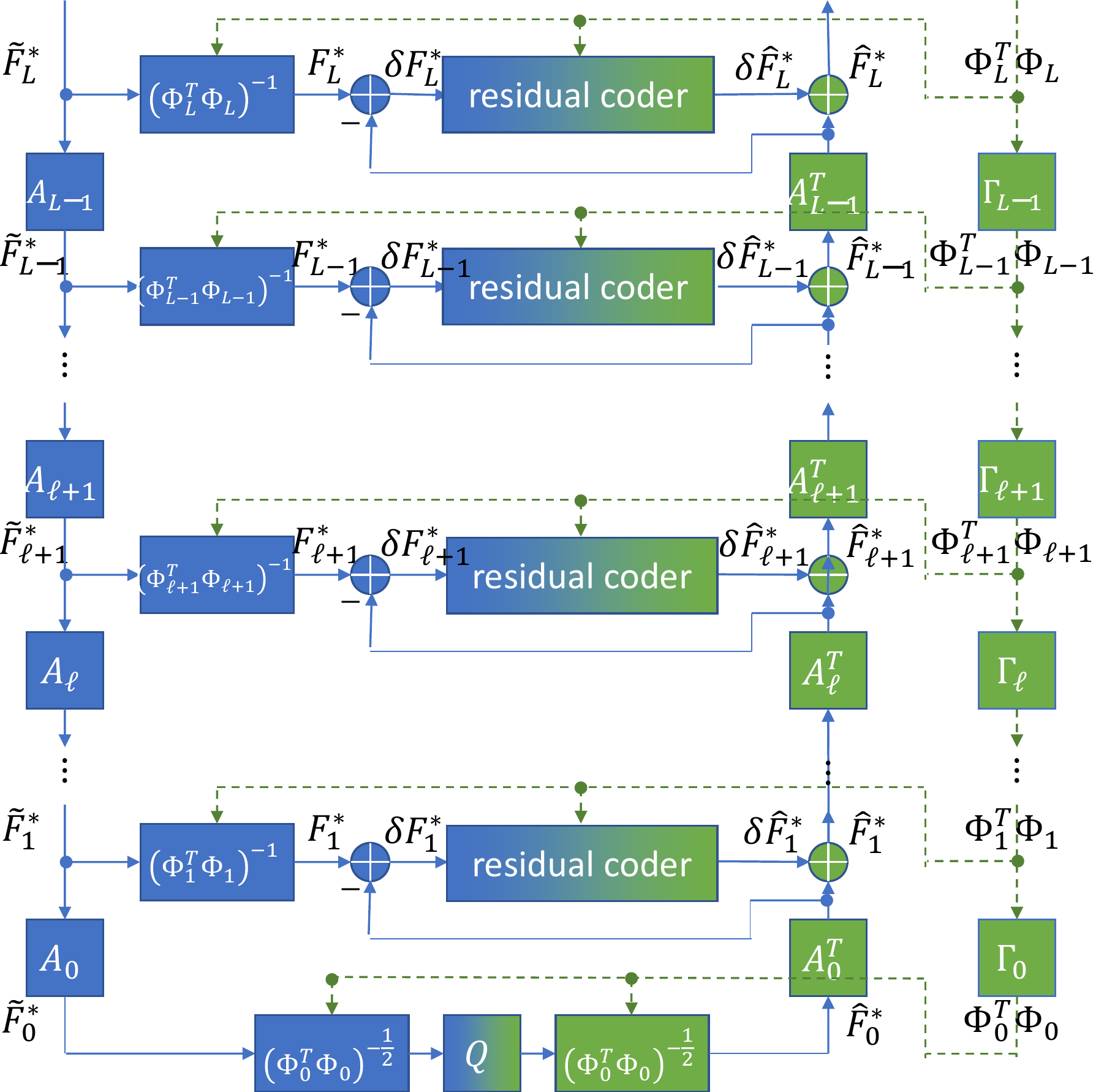}
    \vspace{-0.1in}
    \caption{Multilayer feedforward network implementing point cloud attribute encoder (blue and green) and decoder (green).}
    \label{fig:feedforward_network_block_diagram}
    \vspace*{-3ex}
\end{figure}

\vspace{-0.05in}
We implement the coding framework in the feedforward network shown in Fig.~\ref{fig:feedforward_network_block_diagram}.  The encoding network consists of both blue and green components, while the decoding network consists of only green components.  The encoding network takes as input, at the top left, the point cloud attributes $\tilde{F}_L^*=[\mathbf{y}_i]$ represented as a $N\times r$ tensor, where $N$ is the number of points in the point cloud and $r$ is the number of attribute channels.  (In Sec.~\ref{sec:framework} we assumed $r=1$ for simplicity, but this is easily generalized.)  The attributes $[\mathbf{y}_i]$ may be considered features located at positions $[\mathbf{x}_i]$ in a sparse voxel grid, or they may be considered features located at vertices $[\mathbf{x}_i]$ in a graph.  The network also takes as input, at the top right, the matrix of inner products $\bPhi_L^\top\bPhi_L$ represented as a $N\times 27$ tensor whose $ij$th entry is the inner product
between the basis function $\phi_{L,\mathbf{n}_i}$ located at voxel or vertex $\mathbf{n}_i=\mathbf{x}_i$ and the basis function $\phi_{L,\mathbf{n}_j}$ located at voxel or vertex $\mathbf{n}_j$, the $j$th neighbor in the 27-neighborhood of $\mathbf{n}_i$.  Recall that by construction, this entry is $\delta_{ij}$ at level $L$.

Then from level $\ell=L-1$ to level $\ell=0$, the network computes the $N_\ell\times r$ tensor $\tilde{F}_\ell^*=\mathbf{A}_\ell\tilde{F}_{\ell+1}^*$ by a sparse convolution ($\mathbf{A}_\ell$) of the $N_{\ell+1}\times r$ tensor $\tilde{F}_{\ell+1}^*$ with a $3^3r^2$ space-invariant kernel followed by $2^3$ downsampling, as in (\ref{eqn:analysis_lowpass}).  Similarly, the network computes the $N_\ell\times27$ tensor form of $\bPhi_\ell^\top\bPhi_\ell$ by a sparse convolution ($\mathbf{\Gamma}_\ell$) with a space-invariant $3^3 27^2$ kernel followed by downsampling, as can be derived from (\ref{eqn:PhiTPhi}).  Along the way, the network computes the $N_\ell\times r$ tensor $F_\ell^*=(\bPhi_\ell^\top\bPhi_\ell)^{-1}\tilde{F}_\ell^*$ as in (\ref{eqn:normalization}), by a subnetwork as shown in Fig.~\ref{fig:taylor_expansion_block_diagram}.  The subnetwork implements the operator $\mathbf{X}^{-1}\mathbf{v}$ for the matrix $\mathbf{X}=\bPhi_\ell^\top\bPhi_\ell$ using a truncated Taylor series approximation of $h(x)=x^{-1}$ around $x_0=1/(2\mu)$, where $\mu$ is an upper bound on the eigenvalues of $\mathbf{X}$.

\begin{figure}
    \centering
    \includegraphics[
    height=0.10\textheight, trim=0.5in 4.0in 4.5in 1.125in,clip]{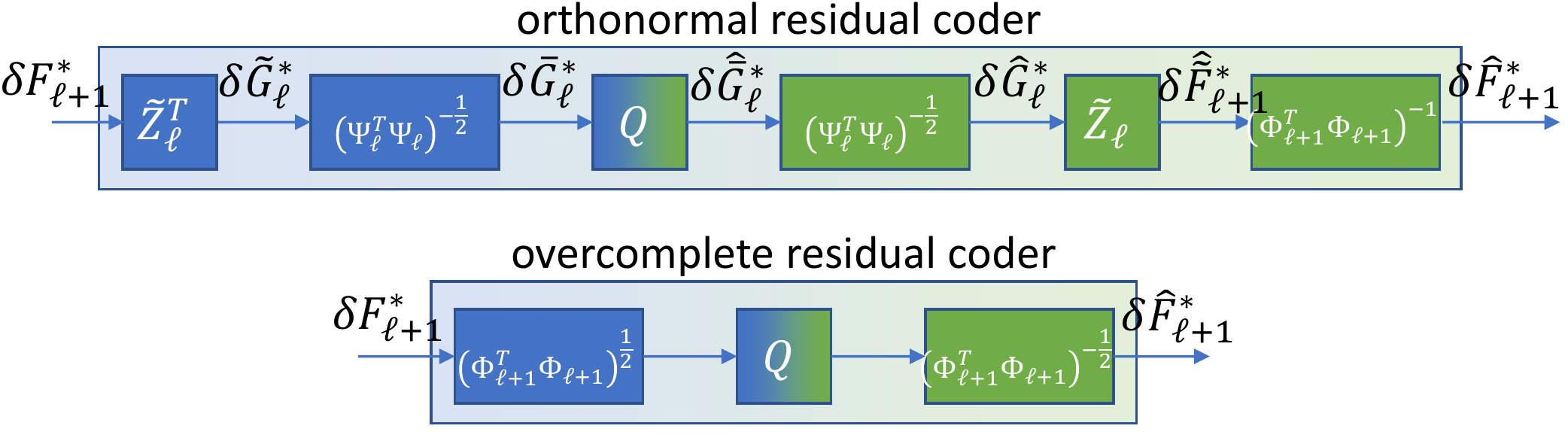}
    \vspace{-0.15in}
    \caption{Multilayer feedforward networks implementing the residual coder subnetwork in Fig.\
    \ref{fig:feedforward_network_block_diagram}.}
    \label{fig:residual_coder_block_diagram}
\end{figure}
\begin{figure}
    \centering
    \includegraphics[
    height=0.06\textheight, trim=0.5in 4.75in 5.0in 1.125in,clip]{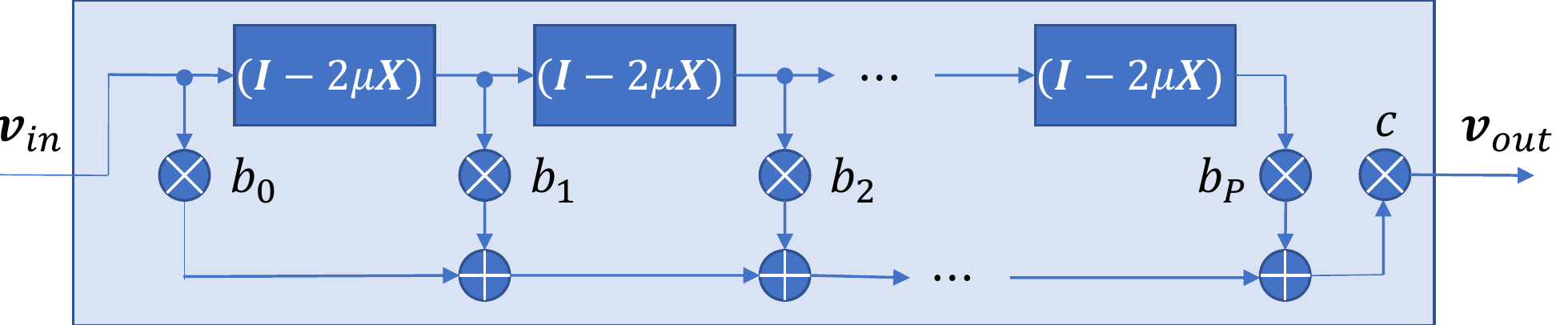}
    \caption{Multilayer feedforward network implementing the subnetworks $h(\mathbf{X})$ in Figs.\ \ref{fig:feedforward_network_block_diagram} and\ \ref{fig:residual_coder_block_diagram}, where $h(x)=x^{-1}$, $x^{-1/2}$, or $x^{1/2}$ and $\mathbf{X}=(\mathbf{\Phi}_\ell^\top\mathbf{\Phi}_\ell)$, $(\mathbf{\Phi}_\ell^\top\mathbf{\Phi}_{\ell+1}^a)$, or $(\mathbf{\Psi}_\ell^\top\mathbf{\Psi}_\ell)$.  Here, $b_0,\ldots,b_P,c$ are coefficients of the $P$th order Taylor expansion of $h(x)$ around $1/(2\mu)$, where $\mu$ is an upper bound on the eigenvalues of $\mathbf{X}$.}
    \label{fig:taylor_expansion_block_diagram}
    \vspace*{-3ex}
\end{figure}

At level $\ell=0$, the features $F_0^*$ are normalized at the encoder by $(\bPhi_0^\top\bPhi_0)^{-1/2}$ as in (\ref{eqn:orthor_normal_equation_F}) and uniformly scalar quantized and entropy coded.  At the decoder (and encoder) they are decoded and re-normalized again by $(\bPhi_0^\top\bPhi_0)^{-1/2}$ as $\hat{F}_0^*$.  Here, the operator $(\bPhi_0^\top\bPhi_0)^{-1/2}$ is implemented by a subnetwork as shown in Fig.~\ref{fig:taylor_expansion_block_diagram}.

Then from level $\ell=0$ to level $\ell=L-1$, the network computes the residual $g_{\ell+1}^*$ represented in the basis $\bPhi_{\ell+2}$ as the $N_{\ell+1}\times r$ tensor $\delta F_{\ell+1}^*=F_{\ell+1}^*-\mathbf{A}_{\ell+1}^\top\hat{F}_\ell^*$ using a sparse transposed convolution followed by $2^3$ upsampling and subtraction, as in (\ref{eqn:delta_F}).  The representation $\delta F_{\ell+1}^*$ is coded by one of the residual coder subnetworks shown in Fig.~\ref{fig:residual_coder_block_diagram}, and decoded as $\delta\hat{F}_{\ell+1}^*$.  
The residual coder subnetworks are again feedforward networks with operators implemented by appropriate Taylor series approximations as in Fig.~\ref{fig:taylor_expansion_block_diagram}.

Note that the feedforward network in Fig.~\ref{fig:taylor_expansion_block_diagram}, which approximates an operator $h(\mathbf{X})$ using a truncated Taylor expansion, can be regarded as a polynomial of a generalized graph Laplacian for a graph with sparse edge weight matrix $\mathbf{X}$.  Thus the feedforward network can be implemented with a graph convolutional network.  
Alternatively the graph convolutions can be regarded as sparse 3D convolutions with a space-varying kernel, whose weights are determined by a hyper-network $\mathbf{\Gamma}_\ell$.  
In this sense, the hyper-network implements an attention mechanism, which uses the geometry provided as a signal in $\bPhi_L^\top\bPhi_L$ to modulate the processing of the attributes.

\vspace*{-0.5ex}
\section{Experiments}
\label{sec:results}
\vspace{-0.05in}
In this paper, to demonstrate the effectiveness of our normalization approximation, we set the weights of $\A_\ell=[a_{ij}^\ell]$ directly without training such that they correspond to volumetric B-splines of order $p=2$ (as described in \eqref{eq:RAHT2}).
We then compare our pre-defined neural network $RAHT(2)$ with $RAHT(1)$ \cite{QueirozC:16,SandriCKQ:19} for attribute representation and compression.
Evaluations are performed on \textit{Longdress} and \textit{Redandblack} datasets that have 10-bit resolution \cite{dEonHMC:2017}.



In Fig.~\ref{fig:recon_PC_samebpp}, we reconstruct the point cloud at a given bit rate to demonstrate the effectiveness of higher order B-splines. It is clear that \textit{piecewise trilinear} basis functions can visually fit the point cloud much better at a given bit rate. With RAHT(2), blocking artifacts are much less visible, since the higher order basis functions guarantee a continuous transition of colors between blocks. 

\begin{figure}
    \centering
    \includegraphics[width=0.48\linewidth,trim=0.5in 0.625in 0.625in 0.875in,clip]{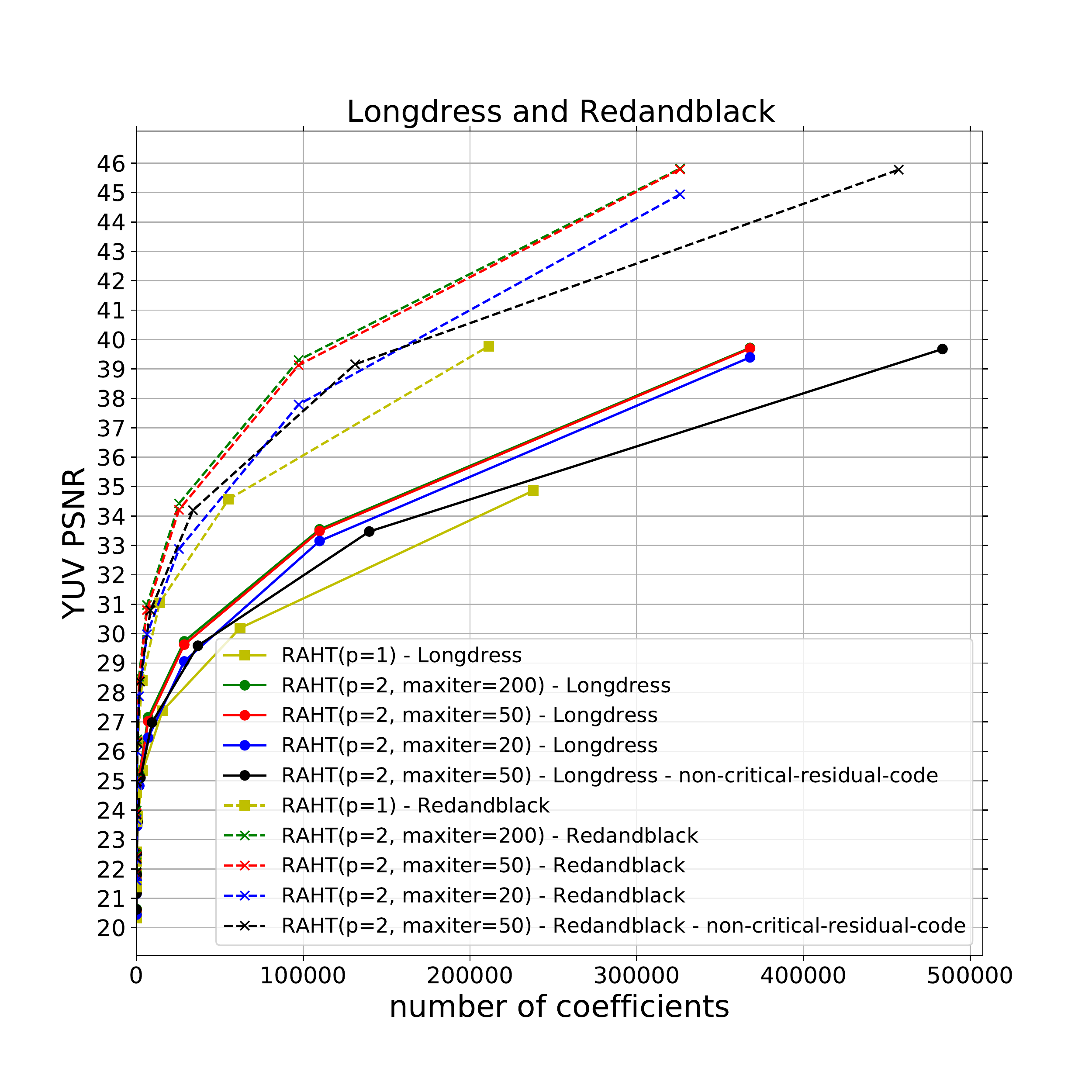}
    \includegraphics[width=0.48\linewidth,trim=0.5in 0.625in 0.625in 0.875in,clip]{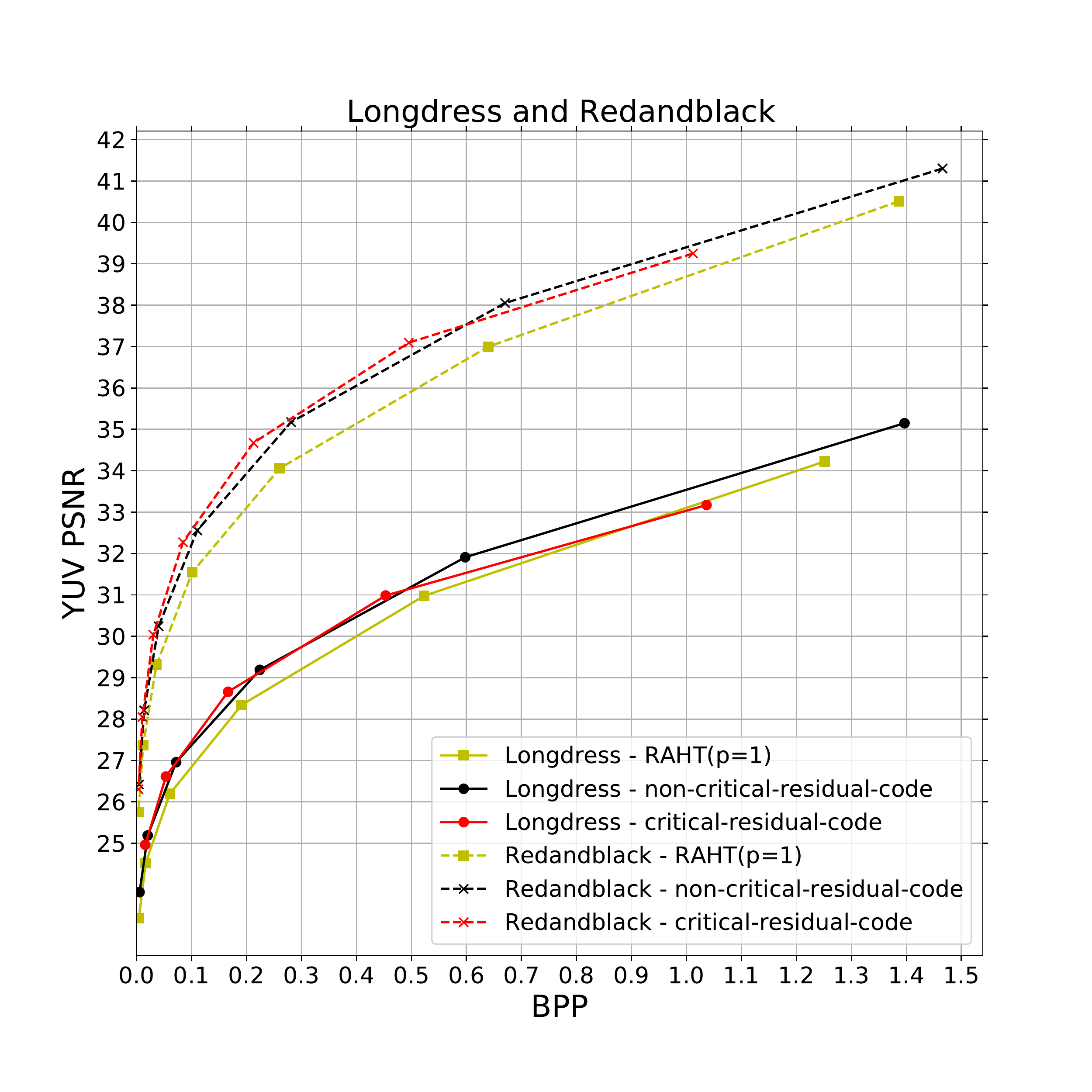}
    \vspace{-0.1in}
    \caption{(left) Energy compaction and (right) Coding gain for both RAHT($p=2$) and RAHT($p=1$), and {\em Longdress} and {\em Redandblack} datasets.}
    \label{fig:energy_compaction_and_code_gain}
    \vspace*{-3ex}
\end{figure}

In Fig.~\ref{fig:energy_compaction_and_code_gain}(left), we show distortion as a function of the number of coefficients corresponding to different levels of detail from 0 to 9. The gap between lines is the improvement in energy compaction. 
We also compare different degrees of the Taylor expansion in our approximation. 
It is clear that RAHT(2) has better energy compaction, achieving up to 2-3 dB over RAHT(1). We also show that overcomplete residual coding has competitive energy compaction compared to RAHT(1) even though it requires more coefficients to express the residual at each level, especially at high levels of detail.

In Fig.~\ref{fig:energy_compaction_and_code_gain}(right), we show YUV PSNR as a function of bits per input voxel. Here, we use uniform scalar quantization followed by adaptive Run-Length Golomb-Rice (RLGR) entropy coding \cite{Malvar:2006} to code the coefficients of each color component (Y, U, V) separately. The plot shows that the improvement of RAHT(2) in distortion for a certain bit rate range can reach 1 dB, or alternatively 20-30\% reduction in bit rate, for both overcomplete and orthonormal residual coding.
At higher bitrates, the performance of orthonormal (critical) residual coding falls off.  We think the reason may have to do with the orthonormalization matrices, which can be poorly conditioned when the point cloud geometry has certain local configurations.  Further investigation is left to future work.  Surprisingly, overcomplete (non-critical) residual coding performs better at these bitrates.  However, at even higher bitrates (not shown), its performance also falls off, tracking its energy compaction performance.  We think the reason is that most of the overcomplete highpass coefficients at high levels of detail, which tend to be zero at low rate, now become non-zero; this puts a heavy penalty on coding performance.

\vspace*{-0.05in}
\section{Conclusion}
\label{sec:conclude}
\vspace{-0.05in}
We investigate 3D point cloud attribute coding using a volumetric approach, where parameters $\theta$ characterizing a target attribute function $f$ are quantized and encoded, so that the decoder can recover $\hat{\theta}$ to evaluate $f_{\hat{\theta}}(\x)$ at known 3D coordinates $\x$.
For signal representation, we define a nested sequence of function spaces, spanned by higher-order B-spline bases, and the coded parameters are the B-spline coefficients, computed via sparse convolution. 
Experiments show that our feedforward network outperforms a previous scheme based on $1$st-order B-spline bases by up to $2$-$3$ dB in PSNR.

\vspace*{-0.05in}
\section{Acknowledgment}
\vspace{-0.05in}
We thank S. J. Hwang and N. Johnston for helpful discussions.

\vfill
\pagebreak

\bibliographystyle{IEEEbib2}
\scriptsize
\bibliography{ref2}

\end{document}